\documentclass[aps,amssymb,amsmath,notitlepage,prb,twocolumn,superscriptaddress]{revtex4-1}
\usepackage{pdfsync}
\usepackage{color}
\usepackage[normalem]{ulem}
\usepackage[caption=false]{subfig}
\usepackage{multirow}
\usepackage[breaklinks=true]{hyperref}
\usepackage[pdftex]{graphicx}

\begin{document}
\title{Adaptively truncated Hilbert space based  impurity solver for  dynamical mean-field theory}
\author{Ara \surname{Go}}
\affiliation{Center for Theoretical Physics of Complex Systems, Institute for Basic Science, Daejeon 34051, Korea}
\affiliation{Department of Physics, Columbia University, New York, New York 10027, USA}
\author{Andrew J. \surname{Millis}}
\affiliation{Department of Physics, Columbia University, New York, New York 10027, USA}
\date{\today}
\begin{abstract}
We present an impurity solver based on adaptively truncated Hilbert spaces. The solver is particularly suitable for dynamical mean-field theory in circumstances where quantum Monte Carlo approaches are ineffective. It exploits the sparsity structure of quantum impurity models, in which the interactions couple only a small subset of the degrees of freedom. We further introduce an adaptive truncation of the particle or hole excited spaces, which enables  computations of Green functions with an accuracy needed to avoid unphysical (sign change of imaginary part) self-energies. The method is benchmarked on the one-dimensional Hubbard model. 
\end{abstract}

\maketitle
\renewcommand{\k}{\mathbf{k}}
\newcommand{\R}{\mathbf{R}}
\newcommand{\gs}{\mathrm{G.S.}}
\newcommand{\gr}{\mathrm{Green}}
\newcommand{\wn}{\omega_n}
\newcommand{\blue}[1]{{\bf\textcolor{blue}{#1}}}
\newcommand{\red}[1]{{\bf\textcolor{red}{#1}}}
\newcommand{\correct}[2]{{\sout{\textcolor{red}{#1}} }{\bf\textcolor{blue}{#2}}}

\section{Introduction}
Dynamical mean-field theory (DMFT)~\cite{Georges1996,Maier2005,Kotliar2006} is widely used to investigate the physics of many-electron systems, and has produced crucial insights into phenomena ranging from the Mott transition to electronically mediated superconductivity. DMFT uses the solution of an auxiliary quantum impurity model (finite number of interacting degrees of freedom coupled to infinite non-interacting bath) to approximate the physics of the full interacting lattice model. However, many important classes of materials lead to impurity models that are too difficult to solve with current methods. For example, systems with partially filled $d$-orbitals and low point group symmetry of the correlated site are not computationally accessible if realistic exchange interactions are included.~\cite{Dang2016}

The key technical issue in implementing DMFT is the solution of the impurity model. Many useful impurity solvers have been proposed and implemented. Continuous-time quantum Monte Carlo~\cite{Gull2011} is widely used in the single-site approximation and can handle relatively large clusters with high symmetry.~\cite{Werner2009, Gull2009} However the method suffers from the fermionic sign problem, which is grievously exacerbated for large clusters or situations of low spatial symmetry (low point symmetry of the correlated ions). Further, the method requires analytical continuation to obtain spectral functions and other experimental observables.~\cite{Jarrell1996, Gunnarsson2010} The numerical renormalization group (NRG)~\cite{Bulla2008} is known to be very accurate for the low energy physics of simple models (e.g. a single impurity with a single band), but extending this method to more general cases has been challenging, although there is recently reported progress for three-orbital models.~\cite{Horvat2016} 

An alternative and also widely used approach to the solution of impurity models is exact diagonalization (ED)~\cite{Caffarel1994}. In ED, one approximates the continuous bath by a finite number of orbitals,~\cite{Koch2008} and solves the resulting finite system exactly. This procedure has many advantages: it excludes uncertainties by fully diagonalizing the given Hamiltonian, it does not require particular assumptions about the symmetry of the problem and it computes real-frequency spectra without involving any nontrivial process (although of course the physically relevant continuous spectra are approximated as a sum of delta peaks). However, the exponential increase of the computational costs required as the system size is increased is a crucial drawback. This exponential growth limits the total number of orbitals (impurity plus bath), $N_s=N_c+N_b$ to $N_s\sim 8$ with a direct diagonalization or $N_s \sim 14$ with iterative schemes such as Lanczos or Davidson method. This is a severe limitation since experience dictates that one needs at least three bath orbitals for each correlated orbital;~\cite{Liebsch2011b} thus models with three correlated orbitals (or four in the case of the single-band Hubbard model, where a high degree of symmetry somewhat reduces the computational burden) are marginally computationally accessible in situations of high point symmetry, but more general cases are computationally inaccessible within the ED method.

As originally formulated, the ED method works with the full Hilbert space of the impurity plus discretized bath system. Quantum impurity models have a unique sparsity structure arising because the interactions exist only on a small subset of `correlated sites', while the bath orbitals are uncorrelated. Recent methodological improvements in effect exploit this sparsity structure. Zgid \textit{et al.}~\cite{Zgid2011,Zgid2012} proposed that one need only deal with a subspace of the full Hilbert space. Zgid and co-workers employed the configuration interaction (CI) approach, familiar from quantum chemistry, to identify a relevant subspace of the full impurity model Hilbert space, and then diagonalized the problem in that subspace. Lin and Demkov further developed the method, noting in particular that CI was most effective when the system had only a few partially occupied orbitals.~\cite{Lin2013} We also developed an active space variant of the CI method, which was applied to the three-band copper oxide model.~\cite{Go2015} 

Because the interactions act only on a small number of sites, in these CI-inspired methods the dimension of the needed subspace grows only slowly with the number of bath orbitals, permitting study of models with more bath orbitals per correlated orbital. However, the required size of the subspace increases rapidly with the number of correlated orbitals. Further, applications to dynamical mean-field theory require a high-quality approximation to the electron self-energy, and even slight inaccuracies in the calculation of the Green function can lead to unphysical impurity self-energies, with imaginary parts that change sign as a function of frequency.  

Recently, Lu \textit{et al.} proposed an iterative scheme to build the truncated Hilbert space by repeatedly applying the Hamiltonian to a certain initial state,~\cite{Lu2014} diagonalizing the Hamiltonian in the truncated space, and then selecting the highest weight Slater determinants in the lowest eigenstate as new initial states.
These authors also proposed a pole-merging scheme to circumvent the unphysical self-energy problem. The approach of Lu \textit{et al.} has to date been applied to single-impurity models with a small number of correlated orbitals. The number of bath orbitals that can be treated is relatively large, and the combination of the pole-merging scheme and the large number of bath orbitals leads to smooth spectra, but the number of correlated orbitals that can be treated is unknown. 

In this paper, we present a new scheme that, within certain restrictions, overcomes these limitations. We build two truncated Hilbert spaces, one for the computation of the ground state and one for the computation of the Green function, by combining two ideas: particle-hole substitution with respect to the reference states and iterative update followed by truncation. We show that this combination enables access to larger systems (more correlated orbitals) than had been previously accessible, while preserving the accuracy needed for computation of the self-energies. 

The rest of this paper is organized as follows. In Sec.~\ref{sec.method} we describe the impurity solver we improved focusing on how we construct the truncated Hilbert space for both the ground state and the Green function.
In Sec.~\ref{sec.Hubbard}, we benchmark the method on the one- and two-dimensional Hubbard model and show that this method can solve the impurity Hamiltonian with eight spin-degenerate (sixteen total) correlated orbitals.
We present scaling of the computational costs with respect to the number of orbitals in the impurity Hamiltonian in Sec.~\ref{sec.cri}. Section~\ref{sec.dis}  explains how to select the parameters to build a cost-efficient truncated Hilbert space. Finally, we give summary and prospects of this method in Sec.~\ref{sec.summary}.

\section{Methods\label{sec.method}}

\begin{figure*}[htb]
	\includegraphics[type=pdf,ext=.pdf,read=.pdf,width=0.99\linewidth]{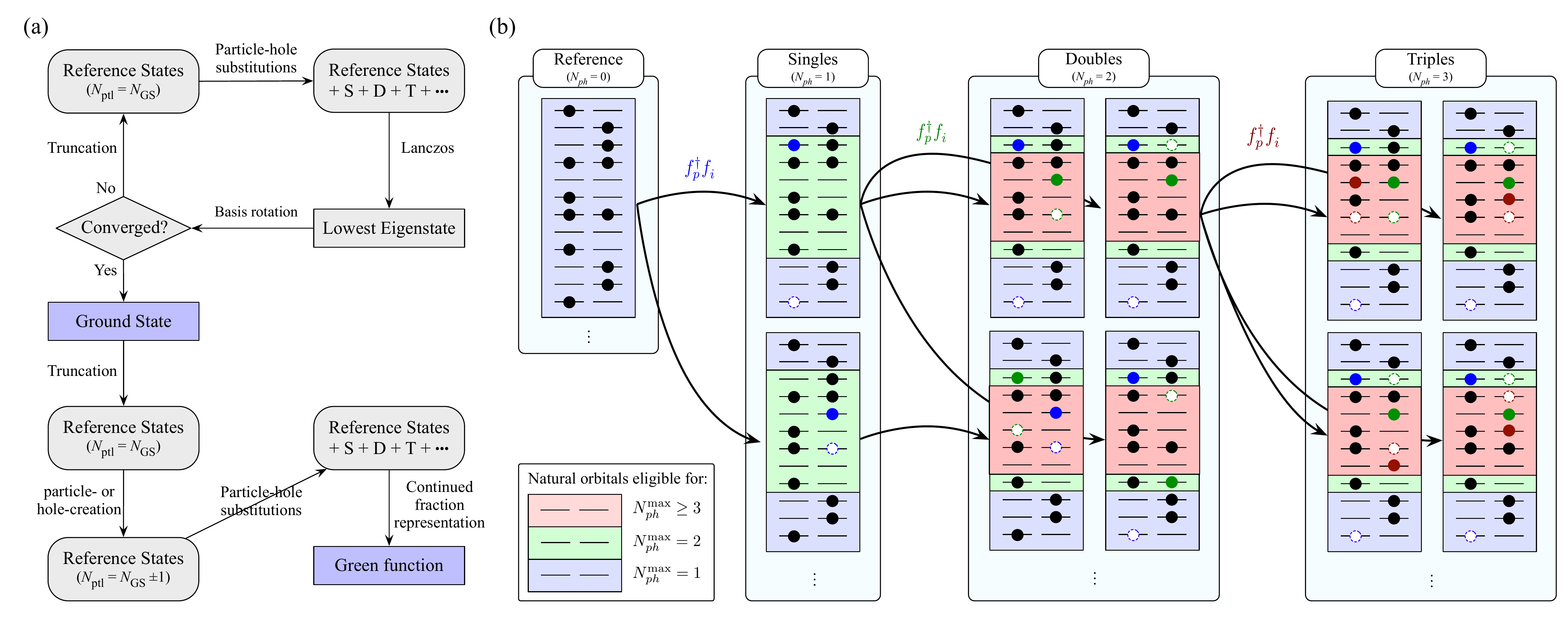}
  \caption{\label{fig.evol}%
  (a) Flow chart of procedure for obtaining the ground state and the Green function: we construct a truncated Hilbert space by applying particle hole substitutions (PHSs) to a set of reference states. The lowest eigenstate is computed within the truncated Hilbert space by the modified Lanczos method. The $N_\mathrm{ref}$ (typically $\mathcal{O}(10^3)$) Slater determinants (SDs) with the highest weight in the ground state are selected as new reference states for next iteration. This procedure is repeated until convergence is reached. Reference states for the Green function calculation are obtained by applying a creation or annihilation operator to the $N_\mathrm{ref}$ highest weight SDs in the final ground state, and the Green function space is again built by adding PHS to these states. The Green function is then computed by the continued fraction representation in this subspace. (b) Selected examples of the generation of SDs by making PHS on reference states. Starting from a set of reference states (one is shown in the leftmost panel), additional SDs are produced by repeatedly applying the PHS to each reference state. Empty (filled) circles indicate holes (electrons) created by the particle-hole substitutions. All SDs in $N_{ph}^\mathrm{th}$ order act as seeds to generate SDs for $(N_{ph}+1)^\mathrm{th}$ order but only few selected SDs are shown for simplicity. At order $N_{ph}=1$ all possible particle-hole pairs are created; at higher order $N_{ph}>1$ particle-hole substitutions are performed only for orbitals which are partially occupied in the natural orbital basis (shaded regions, green and red on-line).}
\end{figure*}

\subsection{Overview}
In this section we outline the key aspects of the methods we present. We are interested in the solution of a quantum impurity model containing $N_s=N_c+N_b$ orbitals, with $N_c$ correlated orbitals created by operators $c^\dagger_\nu$ with $\nu=1,\cdots,N_c$ and $N_b$ bath orbitals created by operators $a^\dagger_l$ with $\nu=1,\cdots,N_b$ (throughout this paper we do not explicitly write the spin indices; $\mu$ and $\l$ label spin-orbital states and each orbital has a two-fold spin degeneracy). The Hamiltonian is
\begin{align}
	H =& \sum^{N_c}_{\mu\nu} \hat{E}^{}_{\mu\nu} c^\dagger_{\mu} c^{}_{\nu}
	+ \sum^{N_c}_\mu \sum^{N_b}_{l} ( V^{}_{\mu l} c^\dagger_\mu a^{}_l + \mathrm{H.c.} ) + \sum^{N_b}_l \varepsilon^{}_l a^\dagger_l a^{}_l \nonumber\\
	+& \sum^{N_c}_{\mu\nu\delta\kappa} I_{\mu\nu\delta\kappa} c^\dagger_\mu c^\dagger_\nu c^{}_\kappa c^{}_\delta,
	\label{eq.hamil_imp}
\end{align}
where $E$ is a matrix describing one particle terms in the impurity, $V_{\mu l}$ is the hybridization strength between the $\mu$th impurity orbital and the $l$th bath orbitals, $\varepsilon_l$ is the onsite energy of $l$th bath orbital, and $I_{\mu\nu\delta\kappa}$ is a matrix element of the two-body interactions among the impurity sites. The circumflex over $E$ denotes an $N_c\times N_c$ matrix. 

The Hilbert space of the problem defined by Eq.~(\ref{eq.hamil_imp}) has dimension $4^{N_s}$, which becomes prohibitively large for $N_s \gtrsim 14$. The original ED method~\cite{Caffarel1994} simply diagonalizes the problem in this Hilbert space. More recent papers \cite{Zgid2012,Lin2013,Lu2014,Go2015} observe that much of the full Hilbert space is irrelevant, and use methods inspired by the CI approach of quantum chemistry to identify the relevant subspace in which the impurity model is diagonalized. 

The standard CI approach diagonalizes the Hamiltonian in a subspace constructed of states obtained by creating some number of particle hole excitations on top of one or more reference states. The reference states are chosen \textit{a priori} (for example as solutions of Hartree-Fock equations) and `active space' methods~\cite{Szalay2012} are used to restrict the size of the subspace by considering only subsets of all possible particle-hole excitations. These approaches have substantially extended the range of impurity models that can be studied \cite{Zgid2012,Lin2013,Go2015} but are not fully satisfactory. They do not take full advantage of the sparsity structure (locality of interactions) characteristic of impurity models, and encounter difficulties in computing the Green function at the level needed to ensure that the sign of the imaginary part of the self-energy is always negative, as required by causality. 

The difficulty with the self-energy $\hat{\Sigma}$ has a straightforward origin: it is defined in terms of the full $\hat{G}$ and bare $\hat{G}_0$ Green functions by the Dyson equation,
\begin{align}
	\hat{\Sigma}(\omega) = \hat{G}^{-1}_0(\omega) - \hat{G}^{-1}(\omega),
\end{align}
In all of the discretized bath methods, both $\hat{G}$ and $\hat{G}_0$ are represented as sums of poles, which must combine correctly to yield a physical $\hat{\Sigma}$. If the Hamiltonian is solved exactly, the self-energy is causal. However in the CI-based methods $\hat{G}$ and $\hat{G}_0$ are determined in different ways. $\hat{G}_0$ is given analytically in terms of the impurity model via 
\begin{align}
	[\hat{G}_0^{-1}(\omega)]_{\mu\nu} = \omega - \hat{E}_{\mu\nu} - \sum_l \frac{V^*_{\mu l} V^{}_{\nu l}}{\omega - \varepsilon^{}_l},
\end{align}
while $\hat{G}$ is computed via an approximate procedure involving a subset of the full Hilbert space. If the computation of $\hat{G}$ is insufficiently accurate, a misalignment of poles between $\hat{G}$ and $\hat{G}_0$, or a misestimate of the strengths of the poles in $\hat{G}$ may occur, leading to unphysical behavior of $\hat{\Sigma}$, which causes difficulties in the DMFT convergence. 

In the remainder of this section we first explain our method for defining a restricted subspace in which the ground state is computed and then show how to compute the Green function with sufficient accuracy that the self-energy satisfies needed `causality' properties. The idea for both ground state and Green function/self-energy calculations is to build an appropriate subspace of the full $4^{N_s}$-dimensional Hilbert space of Eq.~(\ref{eq.hamil_imp}).

\subsection{Ground State}
Our approach aims to exploit the sparsity structure of the impurity model, which is that the interactions live only on a small subset of the total number of sites: the bath orbitals are completely noninteracting. The consequence is that most of the Hilbert space is irrelevant, so that the ground state may be found by diagonalizing Eq.~(\ref{eq.hamil_imp}) in an appropriately selected and parametrically smaller subspace of the full Hilbert space. We adopt an iterative procedure which generalizes and extends those of Zgid {\it et al.}~\cite{Zgid2012} and Lu {\it et al}.~\cite{Lu2014}. We use a multireference CI-like procedure to build a variational subspace of Slater Determinants (SDs) by particle-hole excitations from a set of reference states, diagonalize the problem in this subspace, and then choose as new reference states those SDs with highest weight in the ground state, and repeat until the process converges. This algorithm is sketched in panel (a) of Fig.~\ref{fig.evol}. The method requires a choice of initial (``seed'') reference states; we typically use the references of the multi-reference configuration interaction method~\cite{Go2015}. The final results should be independent of the initial choice, and we have verified that this is generically the case. 

Given a set of reference states, the algorithm then generates a family of states in the usual CI manner by applying particle-hole substitutions (PHSs) to each reference state as illustrated in Fig.~\ref{fig.evol}. We classify the states by the number of particle-hole substitutions $N_{ph}$. For a given reference state $|\psi^R \rangle$, the $N_{ph}=1$ manifold is given by all states of the form $|\psi^{p,R}_i \rangle = f^\dagger_p f^{}_i |\psi^R \rangle$, where $f$ is a fermionic operator, either  $c$ or $a$. Each $|\psi^{p,R}_i \rangle$ is then used to generate a set of $N_{ph}=2$ SDs as $|\psi^{pq,R}_{ij} \rangle = f^\dagger_p f^\dagger_q f^{}_i f^{}_j |\psi^R \rangle$, etc. The set of states generated by applying this procedure to all reference states will typically include many duplicates, which must be removed to define the truncated Hilbert space. It is most convenient to remove duplicates after each $N_{ph}^\mathrm{th}$-order PHSs to prevent a larger set of duplications in the next order of SDs.

After the truncated Hilbert space is constructed, the lowest eigenstate of the Hamiltonian in the restricted Hilbert space is computed as a linear combination of SDs,
\begin{align}
	| \Psi_{\gs} \rangle =&
	\sum_R \Big[ C^R | \psi^R \rangle
	+ \sum_{pi} C^{p,R}_i | \psi^{p,R}_i \rangle
	\nonumber\\
	+& \sum_{pqij} C^{pq,R}_{ij} | \psi^{pq,R}_{ij} \rangle 
	+ \sum_{pqrijk} C^{pqr,R}_{ijk} | \psi^{pqr,R}_{ijk} \rangle
	+ \cdots \Big],
	\label{eq.gs}
\end{align}
where $R$ is the reference state index and subscripts $i$, $j$, and $k$ (superscript $p$, $q$, and $r$) run over the filled (empty) orbital in the given reference state. We used the modified Lanczos method~\cite{Gagliano1986}  to compute the lowest eigenstate in this work, but the matrices are typically not large so many diagonalization methods,  for example Davidson,~\cite{Davidson1975}  standard Lanczos,~\cite{Lanczos1950} or even direct diagonalization (if possible) may be used. The magnitude of the coefficients $C$ indicate the importance of the corresponding SD $|\psi\rangle$ is in the ground state. We choose the $N^\mathrm{SD}_{\mathrm{seed}}$ with the largest $C$ to be the new reference states for the next iteration. The other SDs are abandoned and they are not necessarily included in the Hilbert space in the next iteration unless they are regenerated by PHSs from the new reference states. This closes the loop in Fig.~\ref{fig.evol}(a) to obtain the ground state.

The method described here combines two important ideas from previous methods. In conventional multi-reference configuration interaction methods~\cite{Zgid2012}, the Hilbert space is constructed from a fixed reference set; here we iteratively update the reference states. In the approach of Lu {\it et al.}~\cite{Lu2014}, the reference set is iteratively updated but no further PHS is applied after a reference set consisting of a certain number of SDs is chosen. The dimension of the effective Hilbert space is determined by both $N^\mathrm{SD}_{\mathrm{seed}}$ and $N_{ph}$. The combination of two ideas, iterative update and PHSs, enables us to obtain, with a minimal number of SD, a solution of the impurity Hamiltonian which is sufficiently accurate for the purposes of DMFT. Moreover, as explained in the following subsection, this approach provides an optimal starting point for the computation of the Green function.

Two refinements of the procedure are important to note. At each iteration we compute the $N_s \times N_s$ single-particle density matrix $D$ from the ground state in the usual way from the ground state $| \Psi_{\mathrm{G.S.}} \rangle$ and use this density matrix to transform the single particle basis to the natural orbital basis~\cite{Zgid2012,Lin2013,Lu2014} defined by $D$. Here we fully diagonalized the density matrix, allowing the correlated orbitals to be mixed with the bath orbitals by the transformation. The next step of the iteration is done in this natural orbital basis. The iteration is continued until the changes both in ground state energy and density matrix eigenvalues become minimal. Also, we apply different order of PHSs to natural orbitals depending on their occupancy. The partition of the orbitals is marked by different colors in Fig.~\ref{fig.evol}. The first PHS is applied to all orbitals in SDs in the reference states ($N_{ph}=0$), but we allow higher orders of PHS to only $2N_c+4$ ($2N_c$) most partially filled orbitals for $N_{ph}=2$ ($N_{ph} \geq 3$). This idea is based on the active space variation of the CI method and further reduces the dimension of the Hilbert space by excluding SDs generated by higher order of PHS in occupied or empty natural orbitals.

We also observe that unlike in molecular systems in quantum chemistry, the impurity Hamiltonian has two clearly distinguished types of orbitals: correlated $N_c$ orbitals and noninteracting $N_b$ orbitals. Adding more bath orbitals is much cheaper than adding correlated orbitals, because the particle hole substitutions need only account for the interactions relating to the correlated orbitals. In quantum chemistry language, the active space of partially filled orbitals has size related to $N_c$, so it is small relative to the total space.

\subsection{\label{sec.green}Green function}

\begin{figure}[htbp]
	\includegraphics[type=pdf,ext=.pdf,read=.pdf,width=0.99\linewidth]{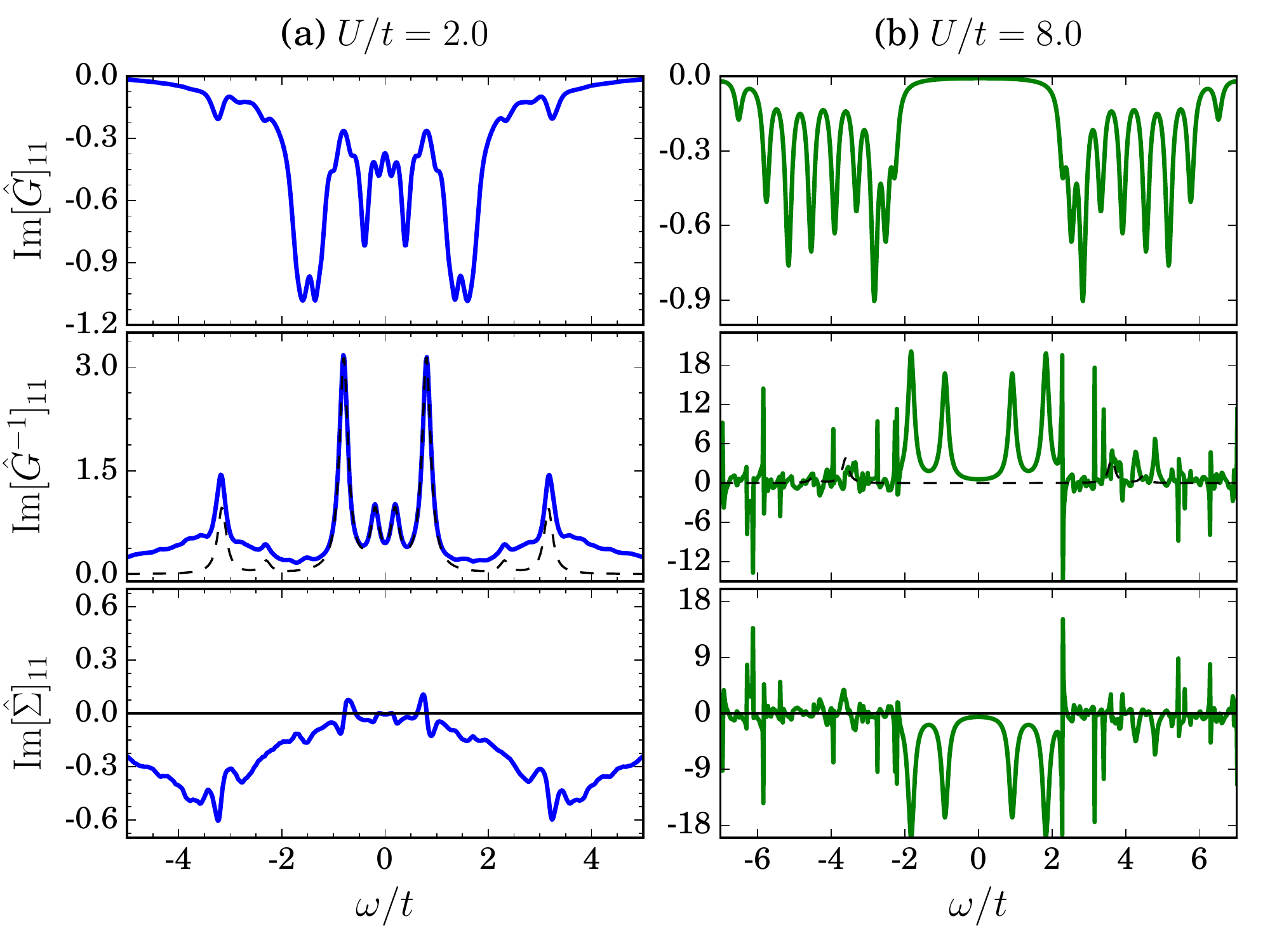}
  \caption{\label{fig.causality}%
  (Top) imaginary part of the on-site term of full Green function computed for the impurity model corresponding to the eight-site cluster dynamical mean-field approximation to the one-dimensional Hubbard model with 16 spin-degenerate bath orbitals at half filling for the $U$ values shown. In this calculation the DMFT solution was converged with respect to basis set size but the Green function calculation was performed using a smaller basis set. (Middle) On-site components of inverse Green function (color on-line) along with (dashed, black on-line) inverse of the noninteracting Green function. (Bottom) Self-energy computed from Dyson equation and inverse of non-interacting Green function.}
\end{figure}

As discussed in the introduction to this section, the need for an accurate self-energy for DMFT computations imposes stringent requirements on the quality of the computed Green function. Figure~\ref{fig.causality} shows examples of the difficulties that can arise, even for apparently high-quality Green functions. The top panels show the imaginary part of the on-site component of the electron Green function, computed from a cluster dynamical mean-field (CDMFT) solution of the half-filled one-dimensional Hubbard model for two values of the interaction parameter. In both cases the Green function is fully causal and has a very reasonable form (note that for the small $U$ case the theoretically confirmed small gap insulating behavior is captured at this level of dynamical mean-field approximation but is obscured by the broadening used to construct the local density of states). The bottom panels show the imaginary parts of the on-site term of self-energy; we see that the imaginary part changes sign. Comparison to the on-site term of the inverse Green function shown in the middle panels reveals that in the $U/t=2.0$ case the unphysical sign change is the result of a small error in the amplitude of the pole at $\omega/t\approx\pm 1$ so that the difference between $\hat{G}_0^{-1}$ and $\hat{G}^{-1}$ is not quite correct. In this figure we also see that the poles in $\hat{G}_0^{-1}$ and $\hat{G}^{-1}$ at $\omega/t\approx \pm 3$ do not quite line up. This misalignment produces unphysical structure in $\hat{\Sigma}$, but in this case it is not large enough to cause a sign change in the imaginary part. The $U/t=8.0$ case reveals an additional difficulty: small inaccuracies in the computed $\hat{G}$ can lead to unphysical behavior in $\hat{G}^{-1}$, arising because in the CDMFT method $\hat{G}$ is a matrix with off-diagonal components, and the poles in the different entries in this matrix must combine correctly to lead to correct behavior in $\hat{G}^{-1}$; the errors in $\hat{G}^{-1}$ are seen to lead to strong causality violation in $\hat{\Sigma}$.

Computing the Green function requires acting on the $N$-particle ground state with an electron creation or annihilation operator and then propagating the resulting state. Implementing this propagation with acceptable computational effort requires constructing a reduced basis set (which we call the Green function space) for the $(N\pm1)$-particle Hilbert space. The standard approach constructs the Green function space by applying a particle or a hole creation operator to each SD used in the ground state.~\cite{Zgid2012, Lin2013, Lu2014, Go2015} In this case the only way to improve the accuracy of the impurity Green function is to increase the number of SDs in the ground state sector itself so that their particle- or hole-excitations produce more SDs in the Green function sectors. Unfortunately in this approach the ground state sector may become excessively large before a sufficiently accurate $\hat{G}$ is obtained. 

The computation of the Green function involves repeatedly applying the Hamiltonian to states in the Green function space. These applications generate particle-hole excitations, increasing the $N_{ph}$. Hamiltonian matrix elements that take the system outside of the space defined from the ground state sector are dropped. The effect of the dropped contributions depends on the weight of the SD from which it originated, because the coefficient persists in the multiplication of the Hamiltonian; SDs produced by adding higher order PHS to an important SD may be more important than states produced by adding lower order PHS to a SD with small ground state weight. This suggests that the desired Hilbert space for the Green function is a set of SDs which minimize the importance of dropped states under as many actions of $H$ as possible, in other words we should choose the set of SDs with PHS which are generated by the Hamiltonian itself.

Motivated by this idea we construct the Green function space as sketched in Fig.~\ref{fig.evol}(a). We build the Green function space by taking the reference states from the ground state calculation, adding a particle or hole, and then completing the space by adding all states generated by adding up to $N_{ph}^\gr$ particle hole substitutions in the $N\pm1$ sectors (in practice we find $N_{ph}^\gs=2$ suffices for the ground state while $N_{ph}^\gr \geq 4$ is needed for the Green function). The Green function is then computed using the standard continued fraction representation within the Lanczos basis constructed from this basis set.~\cite{Mori1965, Caffarel1994,Dagotto1994,Capone2007,Liebsch2009,Lu2014} We observe that it is not necessary to use the same number of reference states as in the ground state. In this work, as a first step we used the same number of seed SDs as in the ground state calculation, and included more SDs in the seed set if the resulting self-energy has causality violations. 

The need to treat only PHS relating to the correlated subspace makes possible our efficient construction of a minimal basis for the Green function space. For constructing this space, we found that treating the first PHS as $N_{ph}=1$, not as $N_{ph}=0$ is useful: the first PHS is applied to at most $2N_c+4$ partially filled orbitals and higher orders of PHSs only target $2N_c$. Since the reference states are already reasonably well selected from the ground state, one can obtain accurate Green functions without involving all orbitals by PHSs. This implies that if a good starting point is chosen, the computational costs even for the ground state may be reduced further by excluding empty and occupied orbitals for PHSs, but we have not yet explored this scheme for the ground state computations. We refer the readers to Sec.~\ref{sec.cri} for further details.

\section{Benchmarking the method on the Hubbard Model\label{sec.Hubbard}}

\begin{figure}[tbp]
	\includegraphics[width=\linewidth]{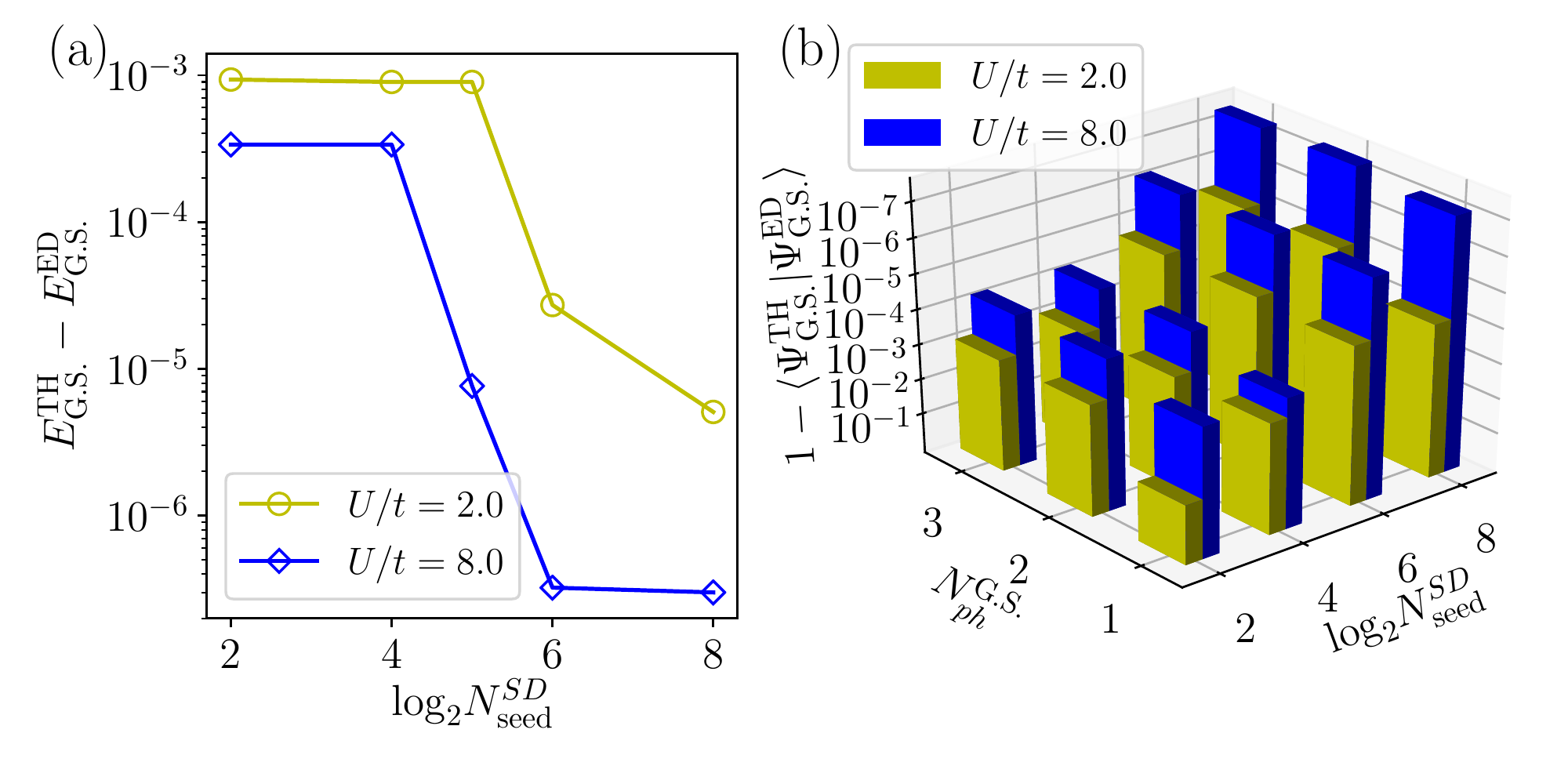}
  \caption{\label{fig.conv_gs}%
  	(a) Error of the ground state energy as a function of the number of seed states with $N^\gs_{ph}=2$. (b) Residual overlap (difference between unity and inner product of the ED ground state and lowest eigenstate from the truncated Hilbert spaces) with various combination of the number of seed states $N^{SD}_\mathrm{seed}$ and the order of PHSs $N^\gs_{ph}$.
  	Semilogarithmic scale is employed to visualize small differences.
  	As $N^{SD}_\mathrm{seed}$ and $N^\gs_{ph}$ increase, the overlap between two wave functions quickly approaches unity.
  	To observe the convergence, the same impurity Hamiltonian with ($N_c$, $N_b$) = (4, 8) was solved by the ED and the new impurity solver using the truncated Hilbert space.
	The bath parameters in the impurity Hamiltonian were taken from the converged DMFT solution for a given value of $U$ in 1D HM.
	See also Table~\ref{tb.size} for detailed information on the Hilbert spaces.
  }
\end{figure}

\begin{figure}[tbp]
	\includegraphics[width=0.99\linewidth]{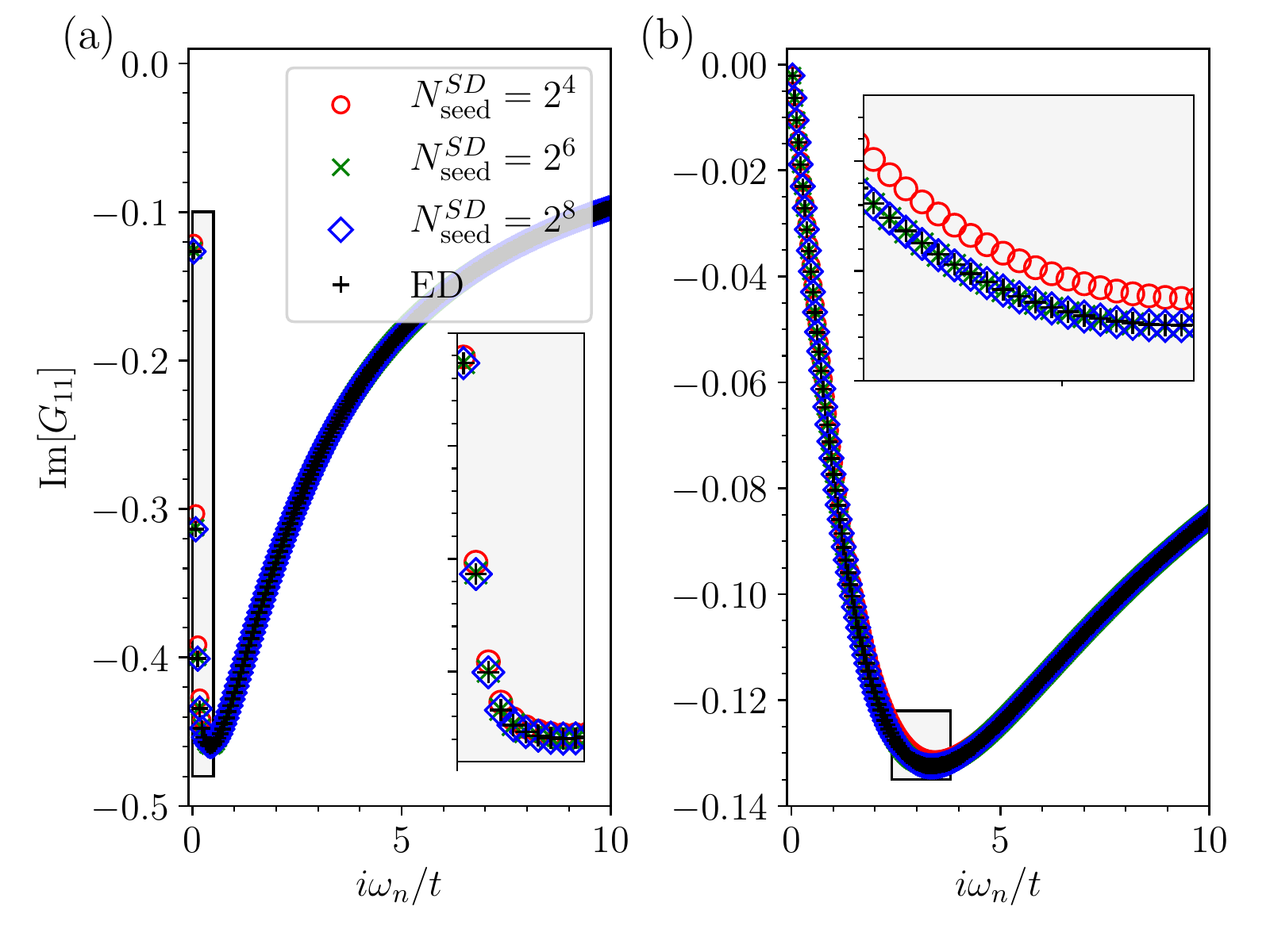}
  \caption{\label{fig.conv_green}%
  	Imaginary part of the impurity Green functions of the impurity Hamiltonian used in Fig.~\ref{fig.conv_gs} for (a) $U/t=2.0$ and (b) $U/t=8.0$.
	We choose $N_{ph}^\gs=2$ and $N_{ph}^\gr=\log_2 N^{SD}_\mathrm{seed}$ to show convergence of the Green function.
	Since the differences are very small, we put insets to emphasize shaded regions where the error is largest.
	See also Table~\ref{tb.size} for detailed information on the Hilbert spaces.
 	As the truncated Hilbert space grows, the Green functions from the truncated Hilbert spaces converge to the ED result rapidly, even before the causality is fully restored.
  }
\end{figure}

We benchmarked our method by using it as an impurity solver for cluster dynamical mean-field theory of the Hubbard model. The Hubbard Hamiltonian reads,
\begin{align}
	H = -t \sum_{\langle i, j \rangle} (c^\dagger_i c^{}_j + \mathrm{H.c}) + U \sum_i n^{}_{i\uparrow} n^{}_{i\downarrow} - \mu \sum_{i} n^{}_{i},
\end{align}
where $i$ and $j$ are the site indices, $t$ is the nearest-neighbor hopping amplitude, $U$ is the local Coulomb interaction between spin up and down electron is the same site, and H.c. indicates Hermitian conjugate. We set the chemical potential $\mu=U/2$ so that the system is half-filled.

In this section we focus on the one-dimensional Hubbard model (1D HM) for which important features of the electron spectral function are exactly calculable from the Bethe ansatz solution.~\cite{Lieb1968,Penc1996,Capone2004,Go2009} and are numerically accessible using the density matrix renormalization group (DMRG).~\cite{Kohno2010} We also present a few preliminary results on the 2D Hubbard model. 

We performed CDMFT calculations for the one-dimensional Hubbard model following the procedure defined in Ref~\onlinecite{Kotliar2001}. The DMFT loop involves putting the computed impurity model self-energy $\hat{\Sigma}$ is into the cellular DMFT (CDMFT) self-consistent equation as
\begin{align}
	G^{-1}_{0,\mathrm{new}}(\omega) = \Big[\sum_\mathbf{k} \hat{G}_\mathrm{latt}(\mathbf{k},\omega)\Big]^{-1} + \hat{\Sigma}(\omega),
\end{align}
where the lattice Green function is obtained by
\begin{align}
	\hat{G}_\mathrm{latt}(k,\omega) = \frac{1}{(\omega + \mu)\hat{1} - \hat{t}(\mathbf{k}) - \hat{\Sigma}(\omega)}.
\end{align}
The Fourier transform of the hopping term $\hat{t}(\mathbf{k})$ defines the lattice structure in the DMFT self-consistent equation and $(\omega+\mu)\hat{1}- \hat{G}^{-1}_{0,\mathrm{new}}(\omega)$ defines the hybridization function, which we fit in terms of a finite number of bath parameters by minimizing the distance function as in the standard ED+DMFT approach,~\cite{Caffarel1994,Go2015}
\begin{align}
	\chi^2 = \sum_n^{N_\mathrm{max}} \sum^{N_c}_{\mu\nu} \big| [\hat{G}^{-1}_{0,\mathrm{new}}(i\omega_n) - \hat{G}^{-1}_0(i\omega_n)]_{\mu\nu} \big|^2,
\end{align}
The fitting is done along the imaginary frequency axis; we chose the required fictitious Matsubara frequencies $i\omega_n=i(2n+1)\pi/\beta$ with $\beta t = 128$ retained $N_\mathrm{max}$=512 frequencies. 

Since the CDMFT breaks translational symmetry, we compute the momentum resolved spectral function by symmetrizing the lattice Green function following the original CDMFT prescription~\cite{Kotliar2001}
\begin{align}
	\rho(k,\omega) = - \frac{1}{\pi} \mathrm{Im} \sum_{\mu\nu} \exp[i (\mu-\nu)] [\hat{G}_\mathrm{latt}(k,\omega + i \eta)]_{\mu\nu},
	\label{eq.pd}
\end{align}
where $\eta=0.1t$ is a Lorentzian-type broadening factor.

\begin{figure}[tbp]
	\includegraphics[width=0.99\linewidth]{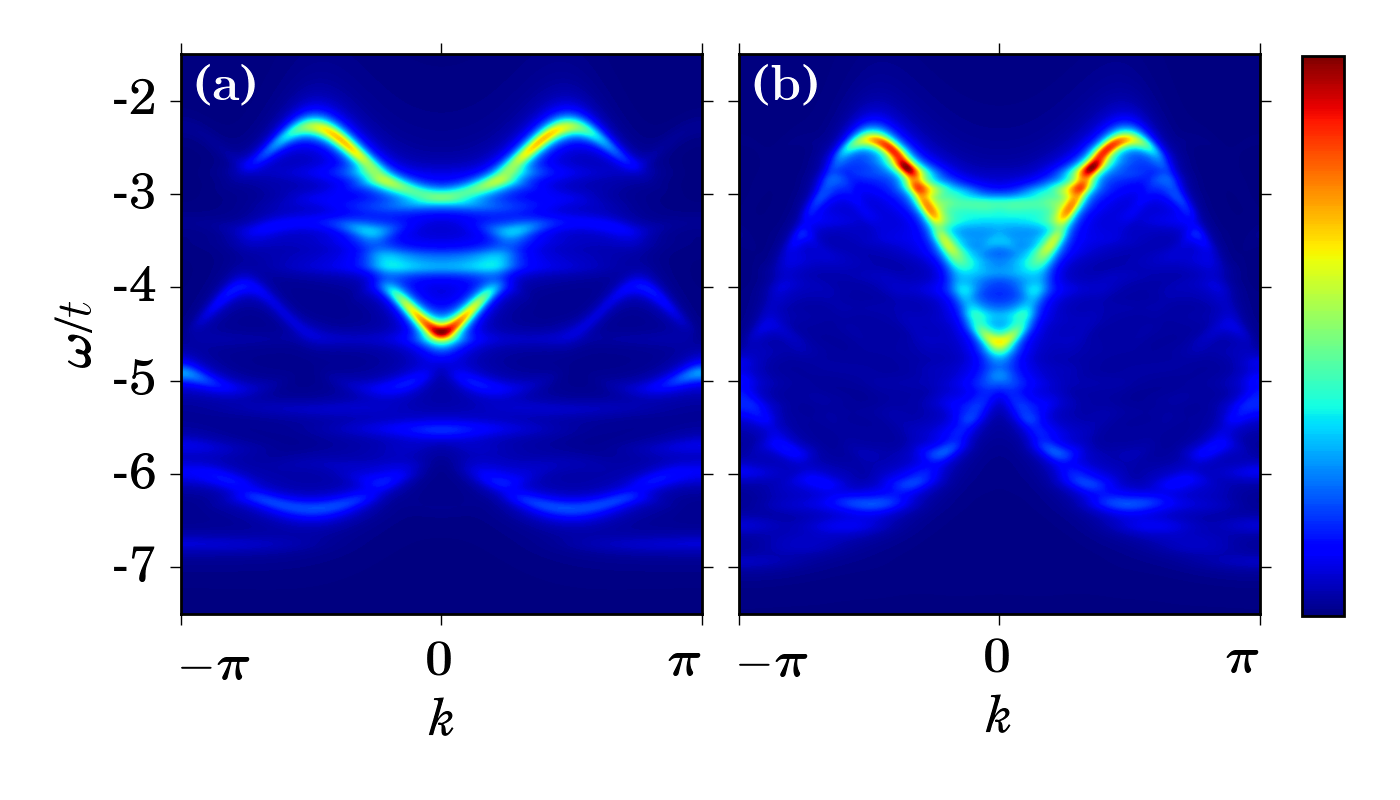}
  \caption{\label{fig.spec}%
 False color representation zero intensity, blue, highest intensity, red, see color bar of the imaginary part of the electron propagator obtained for the one-dimensional Hubbard model at half filling and $U/t=8.0$ with (a) $(N_c,N_b)=(4,8)$ and (b) $(N_c,N_b)=(8,16)$. The same color bar is also used in Fig.~\ref{fig.kdos} and Fig.~\ref{fig.2d_lattice}.
 }
\end{figure}

The level of the CDMFT method is determined by the number of correlated orbitals $N_c$, and our method requires a choice of number of bath orbitals $N_b$.
Figure~\ref{fig.spec} presents the spectral function obtained for two cases: $(N_c, N_b)=(4,8)$ and $(N_c, N_b)=(8,16)$.
The combination $(N_c, N_b)=(4,8)$ corresponds to a Hilbert space dimension $4^{12}=16,777,216$ (use of symmetry reduces the number of states needed in the ED calculation to $853,776$ and is the practical limit of conventional ED calculations for CDMFT studies of the Hubbard model.)
For this case we have verified that our results, obtained with a few thousand states, are numerically equivalent to those from ED.

First, we checked the convergence of the ground state by comparing the lowest energy eigenvalues and by computing the overlap between the ground state wave functions found in our truncated Hilbert space method (TH) and that found by exact diagonalization (ED),
	$\langle \Psi^\mathrm{TH}_\gs | \Psi^\mathrm{ED}_\gs \rangle$.
Figure~\ref{fig.conv_gs} shows that even a very small number of seed states and PHSs reproduces the ground state within numerical accuracy.
The error in the ground state energy and residual overlap decreases exponentially as the number of seed states $N^{SD}_\mathrm{seed}$ increases, while increasing the number of particle hole excitations  $N^\gs_{ph}$ to a value $ \geq 2$ do not improve the accuracy significantly.
We present the convergence of the Green function  in Fig.~\ref{fig.conv_green}.
In this case we have an additional control parameter, the order of PHS to build the Green function space, $N^\gr_{ph}$.
Here we choose $N^\gr_{ph} = \log_2 N^{SD}_\mathrm{seed}$ to show the overall behavior as the Green function space increases.
In fact, for small systems (which are tractable by the ED), any combination with $N^\gr_{ph}\geq 4$ yields sufficiently good accuracy to conduct the DMFT self-consistent calculation on the imaginary frequency.
The size of the truncated Hilbert space required to reproduce the ED results is substantially smaller than the original one.
For more details on the number of SDs we need, see also Table~\ref{tb.size}.

\begin{figure*}[tbp]
	\includegraphics[width=0.99\linewidth]{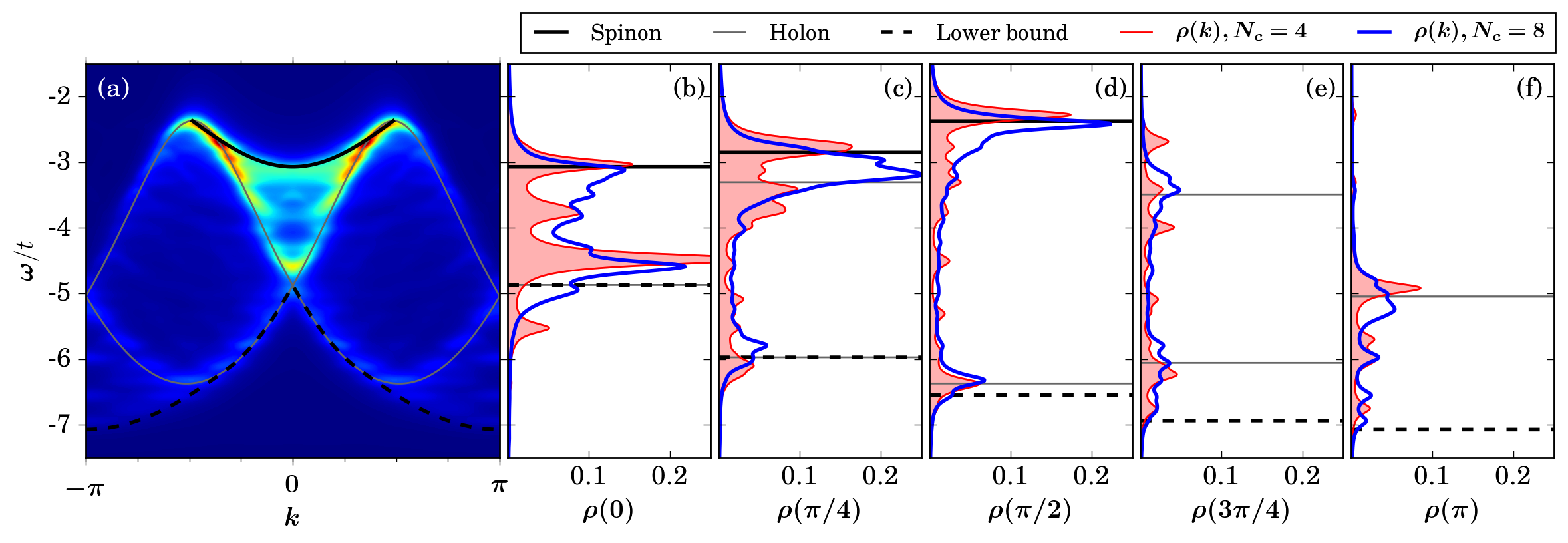}
  \caption{\label{fig.kdos}%
  (a) False color representation (zero intensity, blue, highest intensity, red, see color bar in Fig.~\ref{fig.spec}) of electron spectral function of one-dimensional Hubbard model $U/t=8.0$ from DMFT+truncated Hilbert space approach calculation for $(N_c,N_b)=(8,16)$. Also shown are Bethe ansatz dispersions: spinon branch (black solid line), holon branches (gray solid lines), and lower bound of excitation spectrum (black dashed lines).
  (b)--(f) Blue lines: the $(N_c, N_b)=(8,16)$ electron spectral functions shown in (a) plotted on the $x$-axis against energy on the $y$-axis at $k$-points, $k$=0, $\pi/4$, $\pi/2$, $3\pi/4$, and $\pi$. The spectral functions for the $(N_c, N_b)=(4,8)$ case are given by red lines for comparison. Crossings and lower bounds of the Bethe-ansatz dispersion at the given $k$ are shown by horizontal lines in (b)--(f).
  }
\end{figure*}

The left panel of Fig.~\ref{fig.spec} shows a representative example of a $(N_c, N_b)=(4,8)$ calculation. While (particularly with some a-priori knowledge of the expected spectrum) one can infer a considerable amount of information about the electronic dynamics, it is not clear how which of the structures in the plot are physical and which are artifacts of the approximation. Our method enables us to double the number of correlated orbitals, keeping the ratio of correlated to bath orbitals the same. The right panel of Fig.~\ref{fig.spec} shows results obtained with $(N_c, N_b)=(8,16)$. The additional orbitals lead very obvious improvement in the spectral weights of Fig.~\ref{fig.spec}: the spinon and holon dispersions are clearly distinguished and the triangular spinon-holon continuum is recognizable without any guideline from the Bethe ansatz solution.

Figure~\ref{fig.kdos} presents a more detailed analysis of our results and their relation to previously published results. Panel (a) replots the $(N_c, N_b)=(8,16)$ spectrum from Fig.~\ref{fig.spec}, along with the spinon and holon dispersions and the upper and lower bounds of the spectrum at each $k$, from the Bethe ansatz. Panels (b)-(f) present the spectrum as a function of energy for selected momentum values. We see that the larger cluster size produces a spectrum which is in better agreement with known results, in particular producing larger spectral weights near the energies (shown by solid and dotted lines) where the Bethe ansatz predicts structure. We also see that the large cluster does a better job of concentrating spectral weight in the allowed regions. 

\begin{figure}[tbp]
\includegraphics[width=0.9\linewidth]{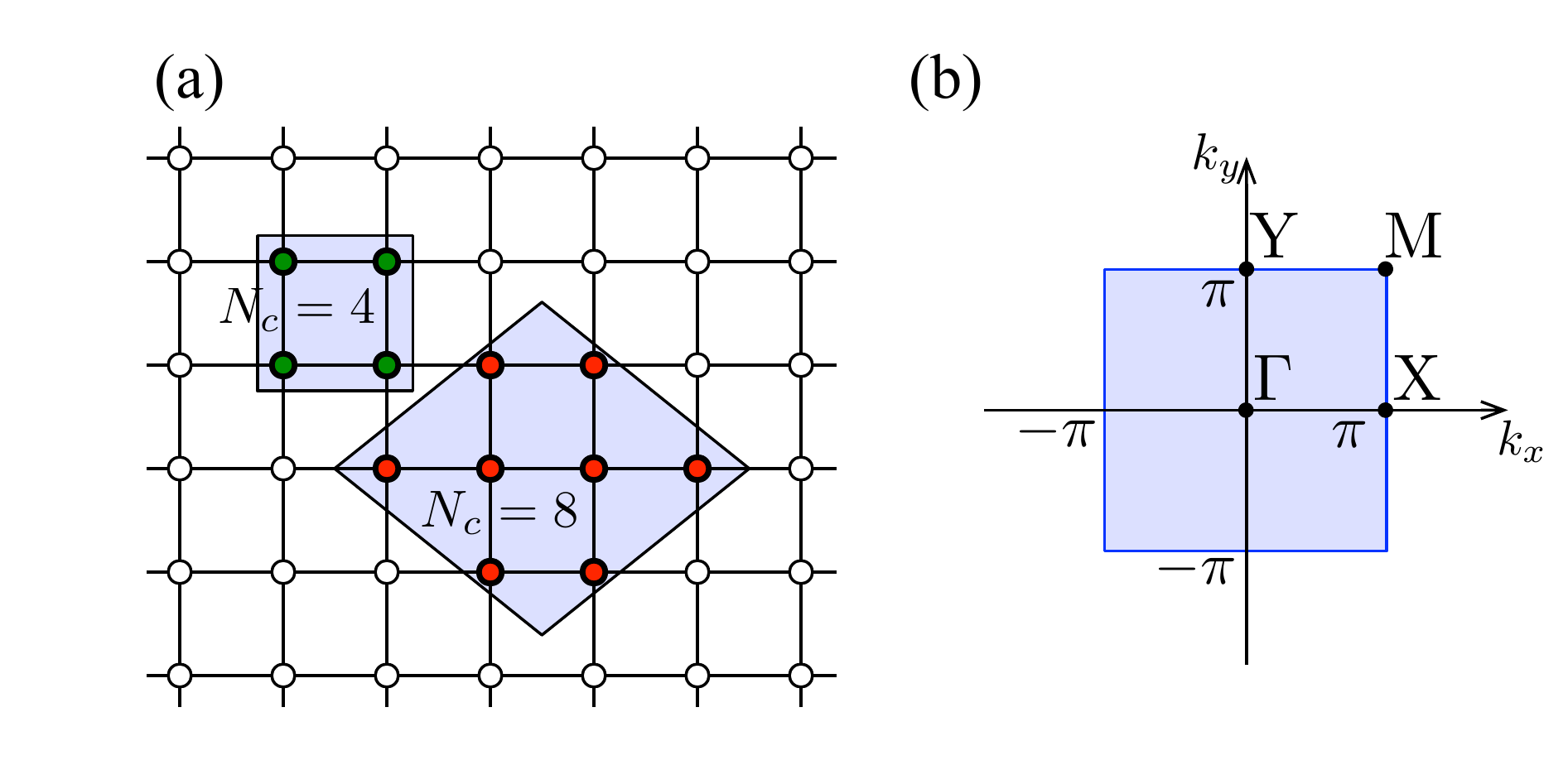}
\includegraphics[width=0.99\linewidth]{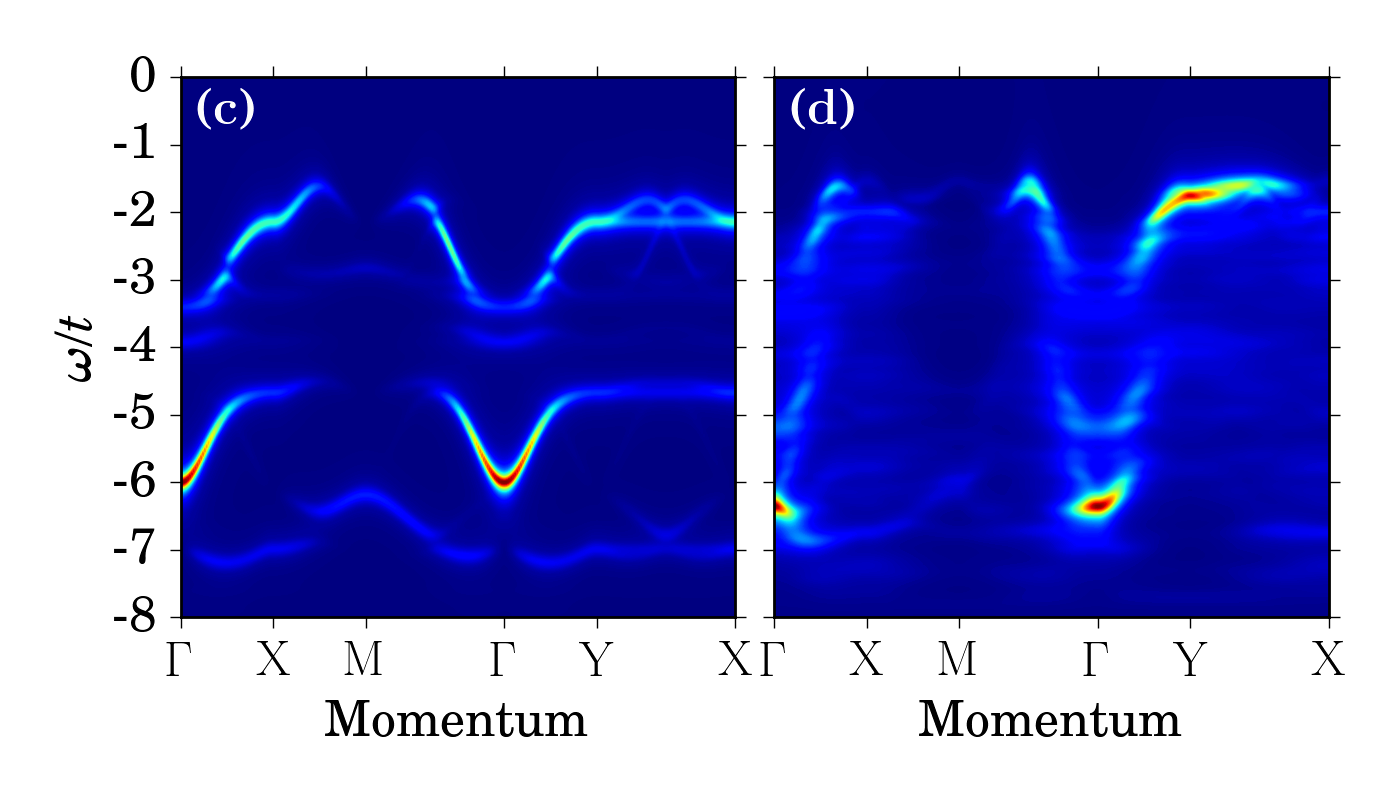}
  \caption{\label{fig.2d_lattice}%
(Top) (a) $(N_c, N_b)=(4,8)$ cluster for the 2D square lattice. The sites in the cluster are highlighted by colors. (b) First Brillouin zone and high-symmetry points of the 2D square lattice. (Bottom) Spectral function of the two-dimensional Hubbard model for $U/t=8.0$ with (c) ($N_c$, $N_b$) = (4, 8) and (d) ($N_c$, $N_b$) = (8, 16).
}
\end{figure}

We also performed CDMFT calculations for the two-dimensional square lattice Hubbard model, using the clusters illustrated in Fig.~\ref{fig.2d_lattice}(a). While $N_c=4$ cluster preserves the $C_4$ point symmetry, there is no $N_c=8$ cluster for CDMFT that is compatible with the $C_4$ point symmetry. We periodize the Green function as we do for the 1D HM in Eq.~(\ref{eq.pd}), to restore translational invariance.

We computed the spectral function from the converged self-energy along the lines connecting high-symmetry points of the Brillouin zone shown in Fig.~\ref{fig.2d_lattice}(b); results are shown in panels (c) and (d) for $(N_c, N_b)=(4,8)$ and $(N_c, N_b)=(8,16)$.
The CDMFT convergence was not particularly more difficult in 2D than in 1D HM, implying that the new impurity solver is not highly sensitive to the dimensionality of the underlying DMFT problem or the connectivity within the cluster, although more SDs were required to recover a causal self-energy.
This result is encouraging, considering that DMFT calculations including eight correlated orbital have been reported only in combination with the continuous-time Monte Carlo~\cite{Werner2009, Gull2009, Sakai2012} or semi-classical approximations.~\cite{Lee2013}
The results given by our new method will provide complementary point of view on this problem.
A detailed analysis of the spectral properties of the two-dimensional Hubbard model obtained by this method will be discussed in a separate paper.
Here we merely remark that the continuous nature of the electron spectral function is much better represented in the larger $N_c$ calculation. 

\section{Convergence Criteria\label{sec.cri}}

\begin{figure}[tbp]
	\includegraphics[type=pdf,ext=.pdf,read=.pdf,width=0.99\linewidth]{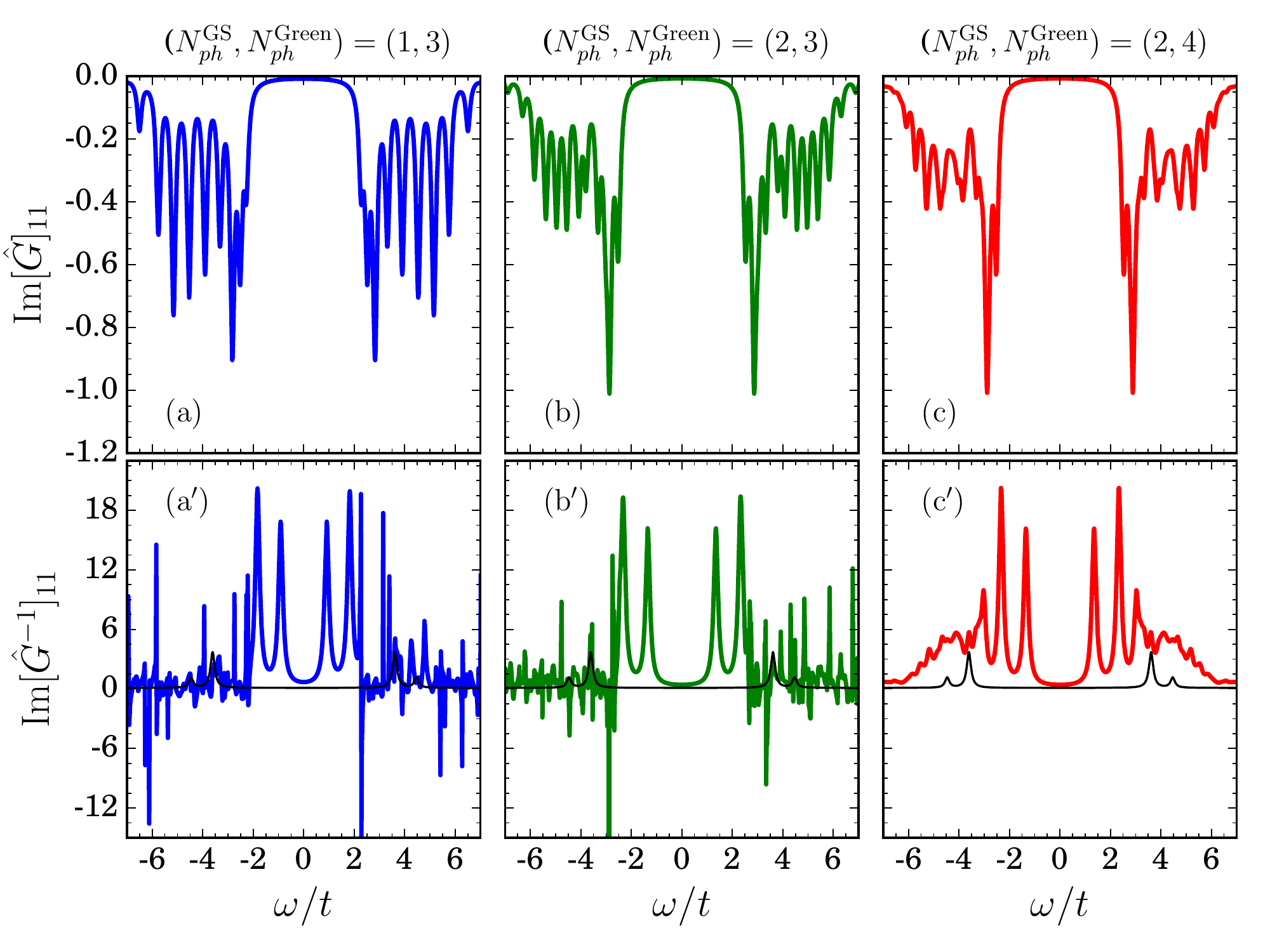}
  \caption{\label{fig.dos}%
  Comparison of the imaginary part of (a)-(c) the Green function and (a$^\prime$)-(c$^\prime$) the inverse Green function from the same impurity Hamiltonian for $(N_c, N_b)=(8,8)$ with different level of accuracy.
  The accuracy is controlled by the order of particle-hole substitutions $N^\mathrm{G.S.}_{ph}$ ($N^\mathrm{Green}_{ph}$) to construct the truncated Hilbert for the ground state (the Green function).
  Only the first diagonal element is shown for clarity but all the other elements show similar behavior for a given frequency.
  All data sets are from the exactly bath parameters which are the converged solution of the DMFT calculation for the 1D HM with $U/t=8.0$.
  While the causality is not broken in the Green function, the inverse Green function ill-behaves.
  The noninteracting inverse Green function is also given in lower panels as black thin solid lines for comparison.
  The corresponding self-energies are given in Fig.~\ref{fig.self}.
  }

\end{figure}

In this section, we discuss more details of our method, focussing on convergence criteria and the trade-offs between computational cost and accuracy of result when increasing the number of SDs by increasing the number of seeds or particle-hole substitutions. First, for problems that are not too large (such as the $(N_c, N_b)=(4,8)$ case discussed above) we have compared the impurity self-energy obtained by the truncated Hilbert space approach to the numerically exact solution from the ED (see Figs.~\ref{fig.conv_gs} and \ref{fig.conv_green}, for example). We found that if the self-energy computed in the truncated Hilbert space is causal at all frequencies on the real-frequency axis, it also agrees to very high precision with the exact ED self-energy on the imaginary frequency, even though the effective Hilbert space is substantially smaller. For larger systems beyond the capability of the ED, we investigated the convergence of the self-energy as the accuracy is improved. We find that in all the cases we have studied the self-energy converges at systems sizes far below the largest Hilbert space size we can study. The rapid convergence of the imaginary self-energy is of particular technical importance, because the DMFT self-consistent equation is solved in terms of the imaginary frequencies, so efficient calculation of the needed self-energy is helpful. 

We next turn to the question of the Hilbert space requirements for obtaining a causal inverse Green function. The three panels of Fig.~\ref{fig.dos} show the evolution of the site-diagonal matrix element of Green function and the inverse Green function computed for the one-dimensional half-filled Hubbard model at $U/t=8.0$ as the size of the ground state and Green function spaces are increased by adding more particle-hole substitutions. The top panels of Figs.~\ref{fig.dos}(a) and (b) show how increasing the number of particle-hole substitutions in the ground state improves the Green function. We find (not shown) that a further increase in the size of the ground state space does not further improve the Green function. The lower panels however show (as also seen in Fig.~\ref{fig.causality}) that even quite reasonable approximations to $\hat{G}$ do not produce adequate approximations to $\hat{G}^{-1}$. Panel (c) then shows that once one has an adequate ground state, increasing the size of the Green function space provides a decisive improvement to the quality of the inverse Green function, even though changes to the Green function are small on the scale shown in the upper panel. 

\begin{figure}[tbp]
	\includegraphics[type=pdf,ext=.pdf,read=.pdf,width=0.99\linewidth]{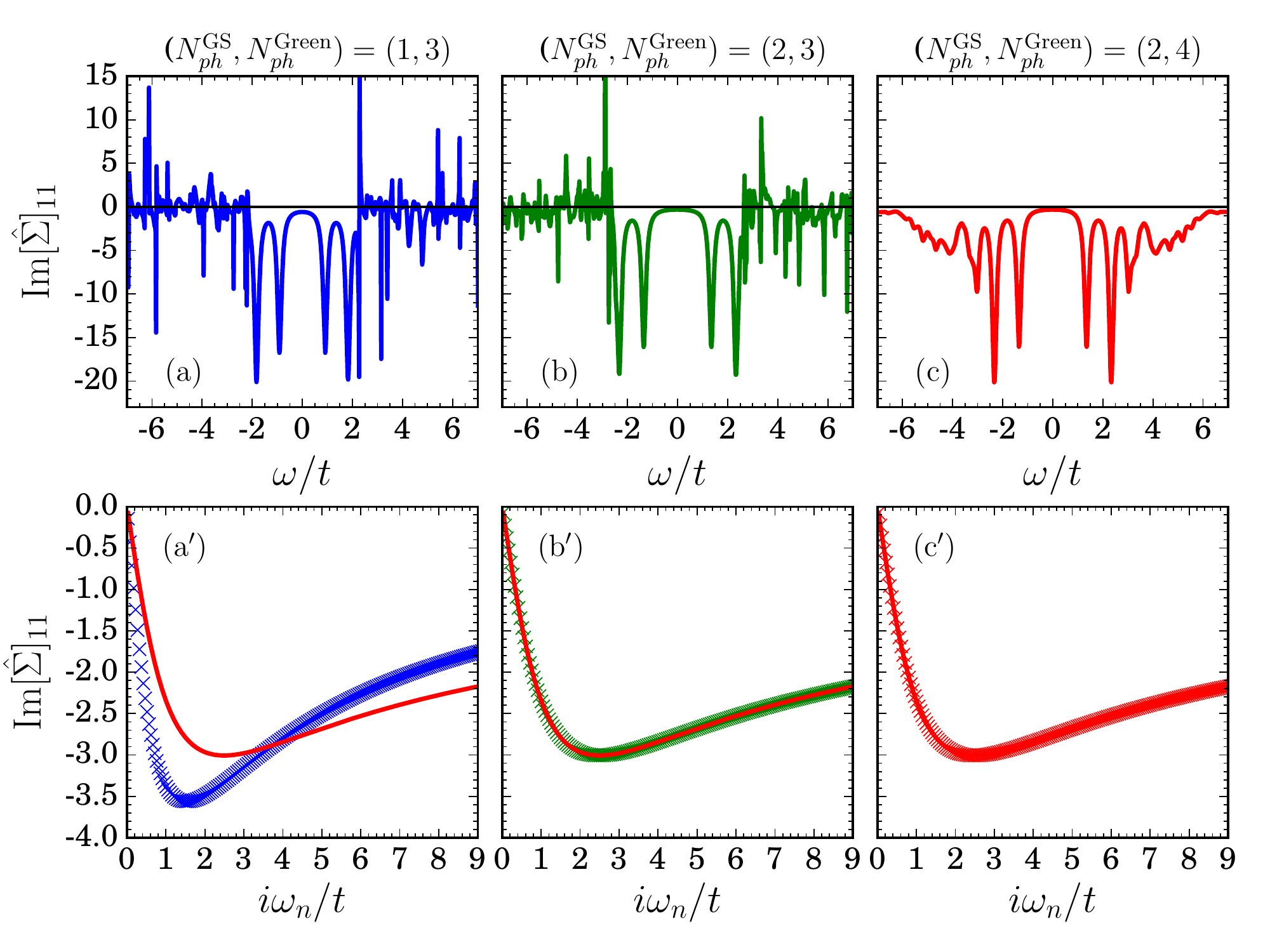}
  \caption{\label{fig.self}%
  Imaginary parts of a diagonal element of self-energy on (a)-(c) the real-frequency axis and (a$^\prime$)-(c$^\prime$) the imaginary frequency axis with the same parameters used in Fig.~\ref{fig.dos}. In panels (a$^\prime$) and (b$^\prime$), the self-energy shown in (c$^\prime$) is presented as a red solid line for comparison.
Even though the imaginary frequency self-energies shown in panels (b$^\prime$) and (c$^\prime$) are indistinguishable on this plot, the real frequency self-energy in (b) shows causality breaking while (c) does not. Causality is restored when the order of $N^\mathrm{Green}_{ph}$ is increased, although the level of accuracy for the ground state is the same in both cases.
  }
\end{figure}

We next turn with Fig.~\ref{fig.self} to the quality of the self-energy, and its relation to the convergence of the DMFT loop. The upper three panels of this figure show the evolution of the self-energy for the three cases shown in Fig.~\ref{fig.dos}. The evolution of the self-energy is very similar to that of the inverse Green function shown in Fig.~\ref{fig.dos}. The lower panels present the Matsubara axis self-energy. We see that already at the intermediate level of accuracy [panel (b$^\prime$)] the Matsubara self-energy is quite accurate (it coincides with the converged result of panel (c$^\prime$), shown as the solid red line)---causality breaking on the level shown in the middle panel is not important for the Matsubara axis computation because the unphysical poles are far enough from the Fermi surface. This indicates that we can conduct the DMFT iteration with lower accuracy at first, and finally improve the accuracy of the Green function after we get the solution. If performing initial computations with lower accuracy, one should carefully check convergence, because our experience is that anomalies near the Fermi level make a difference to the self-energy on the imaginary axis. Specifically, the self-energy evaluated at low Matsubara frequencies is directly affected by the ill-behaving self-energy near zero-frequency on the real axis due to analyticity of the self-energy on the upper-half plane. 

\begin{table*}
	\caption{\label{tb.size}Size chart describing construction of truncated Hilbert spaces large enough to produce causal self-energies in CDMFT calculations of the one-dimensional Hubbard model at density $n=1$ and to recover the causality in the 1D HM with $U/t=2.0$ and $U/t=8.0$. (Left) Number of correlated ($N_c$) and bath $(N_b)$ orbitals, and, for comparison, the number of SDs in the largest symmetry sector of a full exact diagonalization calculation. The middle and right columns show, for the two values of $U$, different cases for the order of particle-hole substitution in the ground $N_{ph}^{\gs}$ and Green function $N_{ph}^{\gr}$ spaces, and the number of `seed' determinants retained, along with the resulting number of SDs in the two spaces. (In principle, the $N^{SD}_\mathrm{seed}$ can be any positive integer but we used powers of 2 starting from 16.) Also shown as $N^{SD}_\mathrm{sig}$ is the number of SDs that have weight larger than $10^{-6}$ in the ground state. (This is not used directly in the program, but shows how the number of important SDs scales as the system size is increased.)  
	}
	\begin{ruledtabular}
		\begin{tabular}{rrr|ccrrrr|ccrrrr}
			\multirow{2}{*}{$N_c$}	& \multirow{2}{*}{$N_b$}	& \multirow{2}{*}{$N^{SD,\mathrm{ED}}_\gs$} &\multicolumn{6}{c|}{$U/t=2.0$} & \multicolumn{6}{c}{$U/t=8.0$} \\
			\cline{4-15}
			&  &   &    				$N^{\gs}_{ph}$	& $N^\gr_{ph}$	& $N^{SD}_\mathrm{seed}$& $N^{SD}_\gs$	& $N^{SD}_\gr$		& $N^{SD}_\mathrm{sig}$  & $N^{\gs}_{ph}$	& $N^\gr_{ph}$	& $N^{SD}_\mathrm{seed}$& $N^{SD}_\gs$	& $N^{SD}_\gr$		& $N_\mathrm{sig}$  \\
			\hline                                                                                                                                  
			2	&8	&63,504			& 2 	& 4	& 16		&828		&132		&14		& 2	& 4	& 16		&652		&82		& 12	\\
			2	&16	&2,363,904,400		& 2 	& 4	& 16		&7,296		&416		&16		& 2	& 4	& 16		&2,914		&180		& 12		\\
			4	&8	&853,776		& 2 	& 4	& 32		&8,271		&1,074		&196		& 2	& 4	& 32		&2,996		&324		& 80		\\
			4	&16	&34,134,779,536		& 2 	& 4	& 32		&33,327		&2,146		&196		& 2	& 4	& 32		&5,412		&888		& 112		\\
			8	&8	&165,636,900		& 2 	& 8	& 256		&841,435	&543,513	&1,484		& 2	& 6	& 128		&220,132	&23,520		& 1,474		\\
			8	&16	&7,312,459,672,336	& 2 	& 8	& 256		&3,444,682	&758,086	&1,546		& 2	& 6	& 256	 	&740,388	&43,120         & 2,136		
		\end{tabular}
	\end{ruledtabular}
\end{table*}

To gain insight into the number of reference SDs and the needed order of particle-hole substitutions we used our method to solve the CDMFT equations of the one-dimensional Hubbard model for various combinations of $N_c$ and $N_b$. We systematically improved the accuracy of the calculation until causality is restored by increasing the order of PHSs $N_{ph}$ and the number of reference states $N^{SD}_\mathrm{seed}$ as shown in Table~\ref{tb.size}. For a given number of seed SDs $N^{SD}_\mathrm{seed}$, we tested various combinations of the particle-hole substitution in the ground state and the Green function spaces. Once those parameters are set, the number of SDs in the Hilbert space is determined automatically by the iterative updates described in Fig.~\ref{fig.evol}. We find that for ground state properties, $N^\gs_{ph}$=2 suffices, but a larger number of PHS is needed for the Green function space. Increasing the PHS order in the ground state is not effective in restoring causality if the order in the Green function space is too small.  

The computational costs are roughly proportional to the square of the number of SDs, however it is important to note that the running time for a single DMFT iteration depends sensitively on the entanglement of the ground state and the quality of the initial states. For example, in our calculations with $(N_c,N_b)=(8,16)$ and $U/t=2.0$ the shortest full DMFT solution took 3 hours with 256 cores to converge while in the most time consuming case we have observed a single DMFT iteration took 26 hours. When the iteration takes a relatively long time, more than 80\% of the running time was spent to compute the ground state.

The method presented here requires a  finite number of bath orbitals, so bath fitting is always an issue.~\cite{Koch2008, Senechal2010, Liebsch2011b,Go2015} More bath orbitals improves the fitting and reproduces the continuous energy spectrum of the effective environment in the DMFT, but also increases the computational cost. In the original ED, the cost to add a bath orbital is not particularly cheaper than that to add a correlated orbital. In this method, however, the $N_b$-dependence of the computational cost is relatively slow, implying that many more bath orbitals can be added than in other methods. Even if we double the number of bath orbitals in the impurity Hamiltonian, the number of most important SDs do not increase drastically. Also we find that the required values of $N_{ph}$ and $N^{SD}_\mathrm{seed}$ are almost independent of $N_b$; therefore we can first perform calculations with smaller $N_b$ to find an optimal parameters, then increase $N_b$ to save time for searching the optimal $N_{ph}$ and $N^{SD}_\mathrm{seed}$.

The favorable scaling of computational burden with $N_b$ arises from the noninteracting nature of the bath orbitals, which are less entangled with other orbitals, so higher order of PHSs involving them do not contribute to the ground state significantly. This scaling grants us more room to invest the limited computational power on treating the correlated orbitals, which still show exponential growth of the costs with $N_c$. Further optimization (implementation of more efficient parallelization and better algorithm to find a reasonable initial SDs) will enable us to include more correlated orbitals than the $8$ studied here.
We are developing an algorithm to build  initial seed states based on an optimal set of SDs with the same $N_c$ but smaller $N_b$, which will significantly decrease the number of iterations required to obtain the ground state. A substantial improvement in computational time is expected because finding the ground state is at present the most time consuming part of the algorithm.
We expect the limit will be around twelve correlated orbitals, where  standard ED starts to suffer from the size of the Hilbert space.

\section{Discussion\label{sec.dis}}
Our method uses several different  control parameters to tune the truncated Hilbert space; finding the optimum combination can be challenging.
In this section, we discuss how to choose the control parameters, $N^\gs_{ph}$, $N^\gr_{ph}$, and $N^{SD}_\mathrm{seed}$ in details.
The first step is to compute a sufficiently accurate ground state. The accuracy of the ground state is controlled by $N^\gs_{ph}$ and $N^{SD}_\mathrm{seed}$.
Since we have found that $N^\gs_{ph}=2$ suffices in the convergence analysis, we can focus on $N^{SD}_\mathrm{seed}$.
Starting from $N^{SD}_\mathrm{seed} = 8$, we obtain the lowest eigenstate of the Hamiltonian in a generated Hilbert space by the operations described in Fig.~\ref{fig.evol}.
Then we double $N^{SD}_\mathrm{seed}$, and repeat the same procedure in the enlarged Hilbert space.
If the newly obtained ground state energy differs less than $10^{-8}$ from the previous one, we proceed to compute the Green function.
Otherwise we keep doubling $N^{SD}_\mathrm{seed}$ until the convergence is reached.

The $N^{SD}_\mathrm{seed}$ is now used to generate SDs in the Green function space.
For this given number of seed states, we again apply PHSs after we add a hole or an electron to each seed state
and we consider convergence with respect to $N^\gr_{ph}$. The real-frequency Green function and self-energy are sums of poles, and small differences in positions and strengths of poles leads to large variations at frequencies near the pole positions, making the formulation of a quantitative frequency-pointwise convergence condition difficult. If $N^\gr_{ph}$ is too small, we find that the imaginary frequency is not causal (imaginary part changes sign) and for large enough $N^\gr_{ph}$ the imaginary part is causal. The imaginary frequency/imaginary time Greens function and self-energy are much better behaved than the real axis, and a frequency pointwise convergence condition is easy to apply and is robust. We find empirically that an $N^\gr_{ph}$ that is large enough to produce a causal self-energy is more than large enough to produce a converged imaginary axis self-energy and Green function and therefore in practice we use the causality of the real axis self-energy as a convergence condition.

The number of seed SDs required to recover causality depends on not only $N_c$ and $N_b$ but also on the Hamiltonian itself. An extreme example is the noninteracting limit, in which the ground state is described by a single SD and only few SDs reproduce the zero self-energy with high accuracy. In the one-dimensional Hubbard model, the most computationally challenging value of interaction strength was found to be $U/t=2.0$, where the interaction strength is the same with the bandwidth and the physics crosses over from itinerant to localized.

\section{Summary and Future Works\label{sec.summary}}
We developed and implemented an impurity solver inspired by the truncated Hilbert space approaches used in earlier work.~\cite{Zgid2012,Lin2013,Lu2014} The algorithm combines the CI idea of the successively applied particle-hole substitutions~\cite{Zgid2012} with an iterative procedure for updating the truncated Hilbert space~\cite{Lu2014} to construct a compact truncated basis that enables study of a much wider range of correlated electron quantum impurity models. By applying the same ideas to the $N\pm1$ particle Green function spaces we eliminated the causality-breaking problem that has plagued previous results. Impurity Hamiltonians with up to eight correlated orbitals and 24 bath orbitals are solvable in this method, and we expect that future optimization will enable treatment of up to twelve correlated orbitals with at least three bath orbitals per correlated orbital. 
We stress that our method is not more efficient than the ED for small systems. Various iterations to build adequate Hilbert spaces take longer time than one required for the ED. The power of this method is favorable scaling enabling treatment of much  larger systems than can be treated in ED, and especially the favorable scaling with number of bath orbitals, which makes the method very useful for DMFT applications.
We also note that this method specifically targets the ground state in constructing the Hilbert space. In principle it is possible extend the target to few lowest excited states, enabling a treatment of nonzero temperatures, but this problem requires further investigation. Extension to nonzero temperatures would make an interesting combination with self-energy functional theory~\cite{Potthoff2003b} or the DMFT calculation for very low temperatures.~\cite{Capone2007}

We demonstrated that our method provides useful and accurate solutions of the cluster dynamical mean-field equations of the Hubbard model. Going beyond previously reported results we showed that the 8-site CDMFT approximation gives a very good account of the electron spectral function of the one-dimensional Hubbard model, clearly revealing the spinon and holon sectors and the range over which continuum excitations exist. Because our method is not expected to be sensitive to the complexity of the impurity Hamiltonian, it can be applied to a wide range of other situations and in particular will be useful for dynamical mean-field studies of multiorbital models, where a severe sign problem limits the applicability of quantum Monte Carlo methods except in very high symmetry situations. Our ability to incorporate many bath orbitals means that we can obtain good approximations to spectral functions, without the need for analytical continuation.

\begin{acknowledgments}
This work was supported by the US National Science Foundation under Grant DMR-01308236. A.G. also acknowledges support by Project IBS-R024-D1 from the Ministry of Science ICT and Future Planning (Republic of Korea).
\end{acknowledgments}

\bibliography{thesis}
\end{document}